\documentclass[twocolumn,amsmath,nofootinbib,amssymb]{revtex4}
\pdfoutput=1
\usepackage{float}
\usepackage{epsfig}
\usepackage{graphicx}
\usepackage{dcolumn}
\usepackage{bm}
\usepackage{color}
\usepackage{tabularx}
\usepackage{cleveref}
\newcommand{\Od}{{\cal O}}

\def\thebiblio#1{
\begin{center}\bf \large References
\end{center}
\list
{[\arabic{enumi}]}{\settowidth\labelwidth{#1.}\leftmargin\labelwidth
 \advance\leftmargin\labelsep
 \usecounter{enumi}}
 \def\newblock{\hskip .11em plus .33em minus -.07em}
 \sloppy
 \sfcode`\.=1000\relax}

\usepackage{color}
\usepackage{verbatim}
\begin{document}
\preprint{}
\title{%
Cosmological perturbations in coherent oscillating scalar field models
}

\author{J. A. R. Cembranos, A.\,L.\,Maroto and S. J. N\'u\~nez Jare\~no}
\address{Departamento de  F\'{\i}sica Te\'orica I, Universidad Complutense de Madrid, E-28040 Madrid, Spain}

\date{\today}

\begin{abstract}

The fact that fast oscillating homogeneous scalar fields behave as  perfect fluids in average and their intrinsic isotropy have made these models very fruitful in cosmology. In this work we will analyse the perturbations dynamics in these theories assuming general power law potentials $V(\phi)=\lambda \vert\phi\vert^{n}/n$. At leading order in the wavenumber expansion, a simple expression for the effective sound speed of perturbations is obtained $c_{\text{eff}}^2 = \omega=(n-2)/(n+2)$ with $\omega$ 
the effective equation of state. We also obtain the first order correction in $k^2/\omega_{\text{eff}}^2$, when the wavenumber $k$ of the perturbations is much smaller than the background oscillation frequency, $\omega_{\text{eff}}$. For the standard massive case we have also analysed general anharmonic contributions to the effective sound speed. These results are reached through a perturbed version of the generalized virial theorem and also studying the exact system both in the super-Hubble limit, deriving the natural ansatz for $\delta\phi$; and for sub-Hubble modes, exploiting Floquet's theorem. 

\end{abstract}

\maketitle

\section{Introduction}

Rapidly evolving coherent scalar fields have been widely studied in cosmology. Their dynamics is not only important during the reheating epoch after inflation, but they can also support periods of accelerated expansion in both the early universe \cite{Damour-1998,Liddle-1998} or at late times  \cite{Liddle-1999,Sahni-2000}. Concerning the dark matter problem, non-thermal candidates like the axion \cite{axions} or other massive scalar \cite{scalars} or pseudoscalar fields \cite{branons} also fall in this class. These models can be interpreted as Bose-Einstein condensates, where the scalar particles occupy the lowest quantum state of the potential \cite{Sin-1994}. Finally, the possibility of ultra-light scalar fields as dark matter candidates has been explored in different works \cite{Peebles-1999,Matos-2000} by tuning appropriately the potential and initial conditions \cite{Matos-2000}.

The general analysis of a homogeneous oscillating scalar field in an expanding universe was performed by Turner in \cite{Turner}. For a power-law potential $V(\phi)=\lambda \vert\phi\vert^{n}/n$, the rapid scalar oscillations around the minimum of such a potential behave as a perfect fluid with an effective equation of state $\omega=(n-2)/(n+2)$. His results can be recovered by means of a generalization of the virial theorem  \cite{Kamionkowski}. Recently, it has been shown that a fast oscillating abelian vector \cite{Cembranos:2012kk}, non-abelian vector \cite{Cembranos:2012ng} or arbitrary spin field \cite{Cembranos:2013cba} will behave in a very similar way. 

The purpose of this work is to analyse the growth of perturbations in these coherent oscillating scalar
theories for arbitrary power law potential. This subject has been 
mainly studied for harmonic potential models that mimic the standard dark matter case
\cite{Sikivie-2009}, as it happens for the axion field \cite{Khlopov-1985}. It has been
proved by using the linear perturbation theory that the axion was equivalent to
CDM for high enough masses \cite{Ratra-1991,Hwang}. However, gravitational instabilities of
oscillations in a harmonic potential are suppressed on small scales \cite{Khlopov-1985,Hu:2000ke}.
This analysis determines the cut-off in the matter power spectrum and its deviations with respect
to the CDM phenomenology. On the other hand, the dynamical stability (ignoring metric perturbations) of general coherent oscillating scalar dark
energy models has been analysed in different works
~\cite{Damour-1998,Sahni-2000,Taruya-1999,Kamionkowski},
even by considering nonlinear evolutions~\cite{Kasuya:2002zs}.
They conclude that potentials supporting accelerated expansion are
generically unstable with respect to the growth of inhomogeneities.

This work is organized as follows: we will briefly review the standard average approach for the background evolution of a scalar field under a power law potential (Section II), as well as set the equations that rule its perturbations (Section III). After the preliminary discussion, we will analyse the well-known case of a massive scalar by means of an adiabatic expansion approach (Section IV). The perturbations evolution of power-law potential models will be studied following the average approach (Section V). Firstly, we will compute the effective sound speed, which is in general the quantity that rules the evolution, using the perturbed version of the generalized virial theorem.  This method allows to extend previous results to an arbitrary power-law potential. Also, exploiting this equation, we will be able to derive a general expression for a possible anharmonic correction in a massive scalar theory. After that (section VI), we will check the validity of the result for the effective sound speed by studying the exact system of equations (non-averaged) in both super-Hubble and sub-Hubble limits. For small wavenumbers, we will show what is the natural ansatz for $\delta \phi$ (subsection A), whereas in the sub-Hubble limit we will study the exact solution of the non expanding equations thanks to Floquet's theorem (subsection B). Finally, we will see that at high $k$, a cut-off is always expected (subsection C).

\section{Background evolution}
We are interested in studying the cosmological 
evolution of a homogeneous scalar field which is rapidly 
oscillating around the potential minimum.
Let us then consider a scalar field theory in cuved
space-time with Lagrangian
\begin{eqnarray}
\mathcal{L} = \frac{1}{2}\; g^{\mu\nu}\partial_\mu  \phi \; \partial_\nu \phi - V(\phi) \;.
\end{eqnarray}
The equation of motion can be written as
\begin{eqnarray}
\frac{1}{\sqrt{-g}}\; \partial_{\mu} \left( \sqrt{-g}\; g^{\mu \nu} \partial_\nu \phi \right) + V'(\phi) = 0\;,
\end{eqnarray}
where $V'$ represents the derivative with respect to its argument.
 By considering a Friedmann-Lema\^itre-Robertson-Walker  
metric in conformal time $\eta$, the equation of motion takes the form
\begin{eqnarray}
\ddot{\phi} + 2 \mathcal{H} \dot{\phi} + V'(\phi) a^2 = 0\;,  \label{Eqphi}
\end{eqnarray}
where $\mathcal{H} = \dot{a}/a$ and $\;\;\dot{}\equiv \partial/\partial \eta\;$.

The corresponding energy-momentum tensor reads
\begin{eqnarray}
T^\mu_\nu = - \delta^\mu_\nu \left( \frac{\dot{\phi}^2}{2 a^2} - V(\phi) \right) + \frac{\dot{\phi}^2}{a^2}\delta^\mu_0 \delta^0_\nu \;.
\end{eqnarray}

Following the same approach as in previous works \cite{Turner}, we are interested in the cosmological evolution generated by the 
effective energy-momentum tensor obtained after
averaging over the fast scalar oscillations.
Thus, we will concentrate on the average Einstein equations given 
by
\begin{eqnarray}
R_{\mu\nu}-\frac{1}{2}g_{\mu\nu}R=8\pi G \langle T_{\mu\nu}\rangle
\label{Einsav}
\end{eqnarray}
In the particular case of a homogeneous scalar field in 
flat RW background, they reduce to 
\begin{eqnarray}
\mathcal{H}^2 = \frac{8\pi Ga^2}{3}\langle\rho\rangle = \frac{8\pi Ga^2}{3}\left\langle  \frac{\dot{\phi}^2}{2 a^2} + V(\phi) \right\rangle\;; \label{Eins00}
\\
2 \dot{\mathcal{H}} + \mathcal{H}^2 =  - 8\pi G a^2\langle p\rangle =- 8\pi G a^2\left\langle  \frac{\dot{\phi}^2}{2 a^2} - V(\phi) \right\rangle\;. \label{Einsii}
\end{eqnarray}

In order to obtain  the average equation of state of the oscillating scalar, we apply a generalization of the virial theorem \cite{Kamionkowski}. Let us consider that the 
typical frequency of the $\phi$ oscillations is $\omega_{\text{eff}}\gg {\cal H}$ and let us calculate the 
average in a certain time interval $T$ such that it is
large compared to the oscillation period but small compared to
the Hubble  time \cite{Cembranos:2012kk}, i.e.,
$\mathcal{H}^{-1}\gg T \gg \omega^{-1}_{\text{eff}}$  . 

We start by calculating the average of the total derivative
given by  
$\partial_0 \left( \dot{\phi}\phi \right) =  \dot{\phi}^2 + \ddot{\phi}  \phi$. Thus, if the $\phi$ oscillations are
bounded
\begin{eqnarray}
 \left \langle \partial_0 \left( \dot{\phi}\phi \right) \right\rangle =\frac{\dot{\phi} \phi\vert_{ t+ T}- \dot{\phi} \phi\vert_{t}}{T} \sim \mathcal{O}\left(\frac{\omega_{\text{eff}} }{T} \phi^2\right) \label{bounded}
\end{eqnarray}
 
Thus, comparing with  $\langle\dot\phi^2\rangle\sim\mathcal{O}(\omega_{\text{eff}}^2\phi^2)$, we see that (\ref{bounded})
is suppressed by a factor $1/(\omega_{\text{eff}} T)$. 
Using (\ref{Eqphi}) and neglecting $\cal{H}$ terms, we can write
\begin{eqnarray}
\left \langle \frac{\dot{\phi}^2+\ddot{\phi} \phi}{2 a^2}  \right \rangle
&=& \left \langle  \frac{\dot{\phi}^2}{2 a^2} - \frac{ V'(\phi) \; \phi}{2} \right \rangle+\Od\left(\frac{{\cal H}}{\omega_\text{eff}}\right) \nonumber \\
&=&\mathcal{O}\left(\frac{1}{\omega_\text{eff} T} \right) . \label{virial}
\end{eqnarray}
The error introduced by neglecting the total derivative can
be reduced by taking large $T$, so that the 
minimum  limit is set by  $\epsilon\equiv {\cal H}/\omega_{\text{eff}}$

By using these equations, we can reach a useful expression for the  the average equation of state:
\begin{eqnarray}
\omega \equiv \frac{\langle p \rangle}{\langle \rho \rangle}=\frac{\left \langle V'(\phi) \; \phi - 2 V(\phi) \right \rangle}{\left \langle V'(\phi) \; \phi + 2 V(\phi) \right \rangle}+\Od(\epsilon)\;.
\label{eos}
\end{eqnarray}
Considering a power-law potential $V(\phi) = \lambda \vert\phi\vert^n/n$, the last expression results
\begin{eqnarray}
\omega = \frac{n - 2}{n + 2}+\Od(\epsilon) \;,
\label{omega}
\end{eqnarray}
(see \cite{Turner} for an alternative discussion). Therefore, from the conservation equation
\begin{eqnarray}
\dot{\langle\rho\rangle} + 3 (1 + \omega ) \mathcal{H} \langle\rho\rangle = 0\;,
\end{eqnarray}
we can show that the evolution of the average energy density is
\begin{eqnarray}
\langle \rho \rangle = \rho_0 \left( \frac{a_0}{a} \right)^{3 (1 + \omega)}\;.
\end{eqnarray}
so that from the Friedmann equation (\ref{Eins00}) we get:
\begin{eqnarray}
 a(\eta) = a_0 \left(\frac{\eta}{\eta_0} \right)^{\frac{2}{1+3\omega}}. \label{a}
\end{eqnarray}

\section{First order perturbations}
Let us consider now a perturbation on the homogeneous evolution studied above for $\phi$:
\begin{equation}
\phi (\eta, \vec{x}) = \phi (\eta)+ \delta \phi (\eta,\vec{x})\;,
\end{equation}
with $\delta\phi$ a small perturbation.

The only sourced metric perturbations are scalars. Therefore we can write
\begin{eqnarray}
ds^2 = a^2(\eta) \left( (1 + 2 \Phi(\eta,\vec{x}))\; d\eta^2 - (1 - 2\Psi(\eta,\vec{x}))\;d\vec{x}^2 \right);\;\;\;
\end{eqnarray}
where we have chosen the longitudinal gauge for the computation.
By taking the Fourier transformation in the spatial coordinates, the equation of motion for the perturbation results
\begin{eqnarray}
\ddot{\delta \phi}_k &+& 2 \mathcal{H} \dot{\delta\phi}_k - 3 \dot{\Psi}_k \dot{\phi} - \dot{\Phi}_k \dot{\phi} \label{Eqdphi}
\\
&+&  (V''(\phi)\; a^2 + k^2) \delta \phi_k  + 2 V'(\phi)\; a^2 \Psi_k =0\;,\nonumber
\end{eqnarray}
and, following the same approach as for the 
background, the average Einstein equations (\ref{Einsav}) read
to first order in perturbations,
\begin{eqnarray}
 \Psi_k &-& \Phi_k = 0\;, \label{PEinsij}
 \\
\nonumber
\\
- 6 \mathcal{H}^2 \Psi_k &-& 6 \mathcal{H}  \dot{\Psi}_k - 2 k^2 \Psi_k =8\pi G a^2 \langle \delta \rho_k\rangle \;, \label{PEins00}
\\
\nonumber
\\
2 \mathcal{H} \Psi_k &+& 2 \dot{\Psi}_k =-i8\pi G a^2  k_i\frac{\langle\delta  T^0_{\;i}\vert_k\rangle}{k^2}\;, \label{PEins0i}
\\
\nonumber
\\
\ddot{\Psi}_k &+& 3 \mathcal{H} \dot{\Psi}_k + \left( \mathcal{H}^2 + 2 \dot{\mathcal{H}}\right) \Psi_k= 4\pi G a^2 \langle\delta p_k\rangle \;.\nonumber \\ \label{PEinsii}
\end{eqnarray}
where the perturbed energy-momentum tensor components read
\begin{eqnarray}
\delta\rho_k=\delta T^0_0\vert_k &=& \frac{\dot{\delta \phi}_k\dot{\phi}}{a^2} - \Psi_k \frac{\dot{\phi}^2}{a^2} + V'(\phi) \delta \phi_k\;;\label{dE00}
\\
\delta p_k \;\delta^i_j=-\delta T^i_j\vert_k &=& \delta^i_j \left(\frac{\dot{\delta \phi}_k\dot{\phi}}{a^2} - \Psi_k \frac{\dot{\phi}^2}{a^2} - V'(\phi) \delta \phi_k \right)\;;\nonumber \label{dEii}
\\
\\
\delta T^0_i\vert_k &=& \frac{-ik_i{\delta \phi}_k\;\dot{\phi}}{a^2}\;.\label{dE0i}
\end{eqnarray}
and the  effective sound speed is:
\begin{eqnarray}
c^2_{\text{eff}}(k)\equiv\frac{\langle \delta p_k\rangle}{\langle \delta \rho_k\rangle}
\end{eqnarray}

In the following section, we will analyse this system for the simple case of a massive scalar. By making an adiabatic expansion, we will obtain the background solution of $\phi$ and compute the averages of Eqs. (\ref{Eins00})-(\ref{Einsii}) explicitly, together with the corresponding
effective sound speed.

\section{Perturbations of a massive scalar}

Homogeneous scalars are widely considered in cosmology, particularly fast oscillating massive scalars are specially relevant since its perturbations mimic those of dust perfect fluids. This fact makes them a good candidate for solving the dark matter problem.

Assuming a quadratic potential ($n=2$) and redefining the field $\tilde{\phi}= a(\eta) \phi$ in (\ref{Eqphi}), the equation of motion turns into
\begin{eqnarray}
\ddot{\phi} + \left( m^2 a^2 - \frac{\ddot{a}}{a}\right)\phi = 0\;.
\end{eqnarray}

Making an adiabatic expansion \cite{Birrell} with $\epsilon =\mathcal{H}/m a \ll 1$, the solution to the leading adiabatic order results
\begin{eqnarray}
\tilde{\phi} &=& \frac{\phi_s}{\sqrt{2 W(\eta)}} \; \sin \left( \int^\eta W \left( \eta ' \right) d\eta ' \right)\label{adphi}
\\
&& + \frac{\phi_c}{\sqrt{2 W (\eta)}} \; \cos \left( \int^\eta W \left( \eta ' \right) d\eta ' \right)\;,\nonumber
\end{eqnarray}
with $\phi_s$ and $\phi_c$ integration constants, and
\begin{eqnarray}
W^2(\eta) \simeq m^2 a^2 \left( 1 + \mathcal{O} \left( \epsilon^2\right) \right)\;.
\end{eqnarray}
Setting the origin of time adequately, the background field reads
\begin{eqnarray}
\phi \left(\eta\right) = \frac{\phi_c}{a^{3/2}} \; \left[\cos\left( \int^\eta m a\left( \eta ' \right) d\eta '\right) + \mathcal{O} \left(\epsilon^2\right)\right]\;.\nonumber\\
\end{eqnarray}

The average energy-momentum tensor is equivalent to a dust perfect fluid (see Eq. (\ref{omega}) or \cite{Turner})
\begin{eqnarray}
\omega = \frac{\left\langle p \right \rangle}{\left\langle \rho \right \rangle} \simeq 0 + \mathcal{O}\left(\epsilon\right)\;.
\end{eqnarray}

The problem of the perturbations of the massive oscillating scalar has been previously studied in \cite{Ratra-1991,Hwang}. In our case, in order to solve the system, we will combine Equations (\ref{PEins00}) and (\ref{PEins0i}), obtaining an equation which together with (\ref{PEinsij}), form an algebraic system for $\Phi_k$ and $\Psi_k$ as a function of $\phi$ and $\delta \phi$.
\begin{eqnarray}
\Psi_k &=& \Phi_k = 8 \pi G \; \frac{\left\langle \dot{\delta\phi}_k \dot{\phi} + 3 \mathcal{H} \delta\phi_k \dot{\phi} + m^2 a^2 \phi \delta\phi_k \right\rangle}{3 \mathcal{H}^2 - 2 k^2}\nonumber
 \\&+&\Od(\epsilon),
\end{eqnarray}
notice that the origin of the error comes from the use of the virial theorem (\ref{virial}) and (\ref{Eins00}) to pass the term $\dot{\phi}^2 \Psi_k$ to the left-hand side on (\ref{PEins00}). This expression gives $\Psi_k$ as a function of the field and
its perturbation, thus, using (\ref{adphi}) we can obtain the solution of (\ref{Eqdphi}) for $\delta\phi_k$ in an adiabatic expansion. 

As we have just seen the background equations of motion have been solved using a WKB approximation (\ref{adphi}) which can be written as an expansion in the so called adiabatic parameter $\epsilon \equiv  \mathcal{H}/m a$. In order to solve the perturbation equations we can similarly use this method. In this case, there is a new scale, $k$, independent of the other two: $\mathcal{H}$ and $m a$. Therefore, we have two different expansion parameters: the same $\epsilon=\mathcal{H}/m a$ as in the background case and the new parameter $k/\mathcal{H}$. In order to simplify the calculations and work with a single parameter, we will assume that the new parameter $k/\mathcal{H}$ is related to 
$\epsilon$  by $k/\mathcal{H}= \Od(\epsilon^\alpha$).
Thus we will work in different regimes by assigning
different values to the exponent $\alpha$.
Thus for example, as we are interested in cosmological perturbations, we typically
expect $k\sim \mathcal{H}$, i.e. both $\mathcal{H}$
and $k$ of the same adiabatic order, or in other words,
$k / \mathcal{H} \sim \epsilon^0$. However, as we 
will see below, this is not the only interesting range in $k$. For example, we will show that when 
$k^2 \sim \mathcal{H} m a$, the behaviour of perturbations
changes. In such a case, $k / \mathcal{H} \sim \epsilon^{-1/2}$. Finally we will also solve the system for scales $k\sim ma$, i.e., 
$k / \mathcal{H} \sim \epsilon^{-1}$. 

\begin{figure*}[t]
\includegraphics[width=0.8\textwidth]{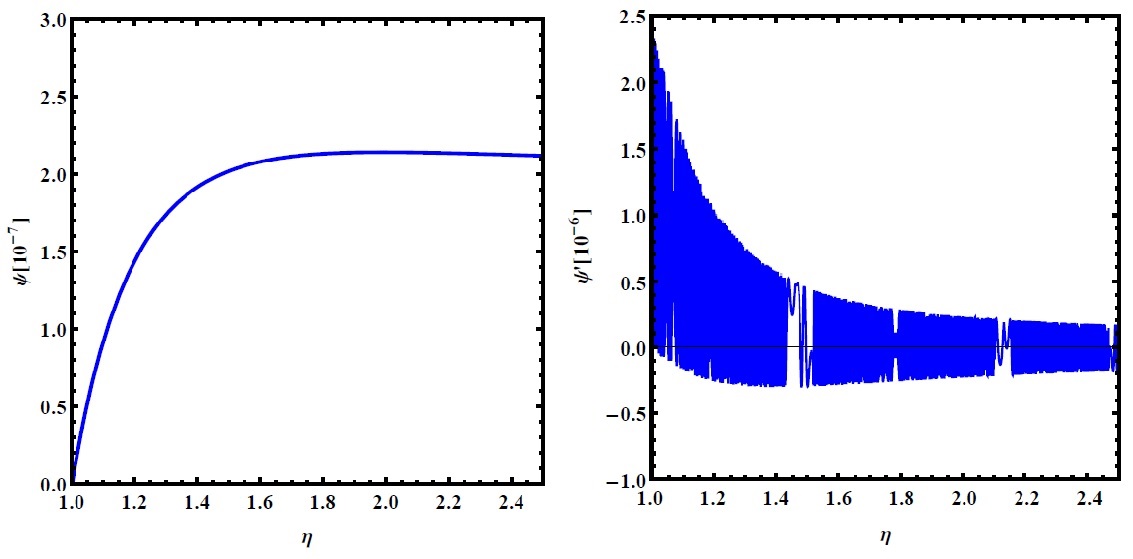}
\includegraphics[width=0.8\textwidth]{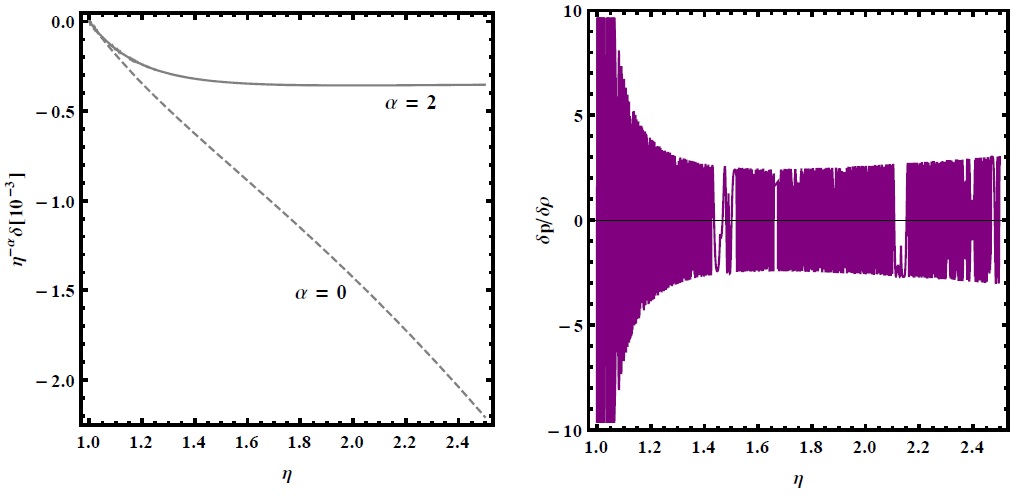}
\caption{\footnotesize{Sub-Hubble mode of a massive scalar field with $k \ll m a$. Those graphics are the numerical solution to the non-averaged Equations (\ref{Eij}$-$\ref{E0i}), with $V(\phi)= m^2 \phi^2/2$. We work in $8\pi G/3 = 1$ and $\eta_{\text{ini}}=1$ units.  In order to calculate a representative mode, we have set $m = 3000$, $k=100$, $\delta\phi(\eta=1)_k= 10^{-6}$, $\dot{\delta\phi}_k(\eta=1) = -10^{-6}$, $a(\eta=1)= 1$. The two plots on the top show that the scalar perturbation of the metric tends to a constant. The energy density contrast (gray) grows as $\eta^2 \sim a$, as expected for the perturbations of a dust-like perfect fluid. Finally, the ratio $\delta p_k/ \delta \rho_k$ (purple) oscillates around zero. The parameter $\alpha$ on the bottom left corner has been introduced to show explicitly the asymptotic behaviour of $\delta$.}}
\label{FigsubH}
\end{figure*}

\subsection{$\alpha \geq 0$}
We will proceed as in the background by assuming an adiabatic ansatz
for the field perturbation,
\begin{eqnarray}
\delta\phi_k (\eta) &=& \delta\phi_s(\eta) \cdot \sin \left( \int^\eta m a(\eta')  d\eta'\right) \label{ansatzdeltaphik}
\\
&+& \delta\phi_c(\eta) \cdot \cos \left( \int^\eta m a(\eta') d\eta'\right)\;,\nonumber
\end{eqnarray}
The  amplitudes can be expanded in the adiabatic parameter,
\begin{eqnarray}
\delta\phi_{s, c} = \delta\phi^{(0)}_{s, c} + \delta\phi^{(1)}_{s, c} +\mathcal{O}(\epsilon^2)\;.
\end{eqnarray}
where $\delta\phi^{(0)}_{s, c}$ and $\delta\phi^{(1)}_{s, c}$
are of adiabatic order $\epsilon^0$ and $\epsilon$ respectively. 
In the standard case $\alpha=0$ ($k\sim {\cal H}$), we have access to super-Hubble and sub-Hubble modes provided  $k^2 \ll m a \mathcal{H}$. From the leading order of the equation of motion (\ref{Eqdphi}), we obtain
\begin{eqnarray}
\delta\phi^{(0)}_{c} = 0\;,
\end{eqnarray}
and once we take this constraint into account, the equation implies
\begin{eqnarray}
\delta\phi^{(1)}_{c} &=& \frac{\mathcal{H}}{24 m a} \left( 3 \left(12 + k^2\eta^2\right) \delta\phi^{(0)}_s + k^2\eta^3 \dot{\delta\phi}^{(0)}_{s}\right),\;\;\;\;
\\
\ddot{\delta\phi}^{(0)}_s &+& 3 \mathcal{H} \dot{\delta\phi}^{(0)}_s = 0 \;.
\end{eqnarray}
By solving this system, we can easily compute the  perturbations to the leading order,
\begin{eqnarray}
\delta\phi^{(0)}_s &=& \frac{C_1}{a^{5/2}} +  C_2  ,\label{deltaphisca}
\\
\Psi_k &=& -\frac{3}{2 \sqrt{2}} C_2 + \frac{1}{5\sqrt{2}} \frac{C_1}{a^{5/2}}  + \mathcal{O}\left(\epsilon\right)\;.\;\;\;
\\
\langle\delta \rho_k\rangle &=& \frac{3 \mathcal{H}^2 + k^2}{\sqrt{2} a^2} C_2
\\
&-& \frac{\sqrt{2}}{15}\frac{-9 \mathcal{H}^2 + 2 k^2}{2 a^2} \frac{C_1}{a^{5/2}} + \mathcal{O}\left( \epsilon\right)\;,\nonumber
\\
\langle\delta p_k\rangle &=&  0 +  \mathcal{O}\left(\epsilon\right)\;,\label{deltapsca}
\\
c_{\text{eff}}^2 &=& 0 +\mathcal{O}(\epsilon)\;.
\end{eqnarray}
As it can be seen the solution mimics the perturbations of dust perfect fluids, i.e. it has vanishing effective sound speed, constant $\Phi_k$ and $\langle\delta \rho_k\rangle/\langle\rho\rangle\propto a$ at
late times. 

 The case $\alpha > 0$ corresponds to super-Hubble modes and has the same solutions (\ref{deltaphisca})-(\ref{deltapsca}) just neglecting $k$ in comparison with $\mathcal{H}$.

\subsection{$\alpha \leq -1/2$}
Let us first consider the  $\alpha = - 1/2$ ($k^2\sim {\cal H} ma$) case. Expanding the equations of motion in $\epsilon$ we obtain the system
\begin{eqnarray}
\delta \phi^{(0)}_s&=& \frac{2 m a}{k^2} \left(\dot{\delta \phi}^{(0)}_c +\frac{ 3 }{2}\mathcal{H} \delta \phi^{(0)}_c \right)\;,
\\
\ddot{\delta \phi}^{(0)}_c &+& 4 \mathcal{H} \dot{\delta \phi}^{(0)}_c + \left( \frac{3}{2}\mathcal{H}^2 + \frac{k^4}{4 m^2 a^2}\right) \delta \phi^{(0)}_c= 0.
\end{eqnarray}
with solution
\begin{eqnarray}
\delta \phi^{(0)}_c &=& \frac{1}{a^{3/2}}\left[ \left( 3 \frac{m a \mathcal{H}}{k^2} C_2 + \left(3 \frac{m^2 a^2 \mathcal{H}^2}{k^4} -1\right) C_1\right) \right.\nonumber
\\
\nonumber
\\
&\times& \cos\left(\frac{k^2}{m a \mathcal{H}}\right) + \left( 3 \frac{m a \mathcal{H}}{k^2} C_1 - \left(3 \frac{m^2 a^2 \mathcal{H}^2}{k^4} \right.\right.\nonumber
\\
\nonumber
\\
&& \left.\left.\left. -1 \right) \; C_2 \right) \;\sin\left(\frac{k^2}{m a \mathcal{H}}\right)\right] \;,
\end{eqnarray}
where $C_1$ and $C_2$ are integration constants. The cosmological perturbations read
\begin{eqnarray}
\Psi_k &=& - \frac{3}{\sqrt{2}} \frac{m a \mathcal{H}}{k^2} \frac{1}{a^{3/2}} \left[ \left( 3 \frac{m a \mathcal{H}}{k^2} C_1 \right.\right.\nonumber
\\
 &+& \left. \left. \left(1 - 3 \frac{m^2 a^2 \mathcal{H}^2}{k^4}\right)C_2\right)\sin\left(\frac{k^2}{m a \mathcal{H}}\right)\right.\nonumber
\\
&+& \left( 3 \frac{m a \mathcal{H}}{k^2} C_2 - \left(1 - 3 \frac{m^2 a^2 \mathcal{H}^2}{k^4}\right)C_1\right)\nonumber
\\
&\cdot& \left. \cos\left(\frac{k^2}{m a \mathcal{H}}\right)\right] + \mathcal{O}(\epsilon)\;,\label{Psiscak}
\\
\nonumber
\\
\left\langle\delta \rho_k\right\rangle &=& - \frac{2}{3} \frac{k^2 \Psi_k}{a^2}+ \mathcal{O}(\epsilon)\;,
\\
\nonumber
\\
c_{\text{eff}}^2 &=& \frac{k^2}{4 m^2 a^2} + \mathcal{O}(\epsilon)\;.\label{Ceffscak}
\end{eqnarray}

In this case $\Phi_k$ oscillates slowly compared to $\omega_{\text{eff}}$ with a decaying amplitude at early times and is constant at late times. On the other hand $\langle\delta \rho_k\rangle/\langle\rho\rangle$ also oscillates with decaying amplitude at early times
and grows as $a$ at late times thus matching the dusty behaviour. 

For modes with $-1/2> \alpha> -1$, the expressions are equivalent to (\ref{Psiscak})-(\ref{Ceffscak}) neglecting the factors that are a power of $m a \mathcal{H}/k$.

Finally, for modes with $\alpha\leq -1$, $\delta \phi$ oscillates with a higher frequency than $\phi$ and all perturbations go to zero in average. Consequently, a cut-off is expected as we consider larger-$k$ modes.
\begin{figure}[h]
\includegraphics[width= 0.85\columnwidth]{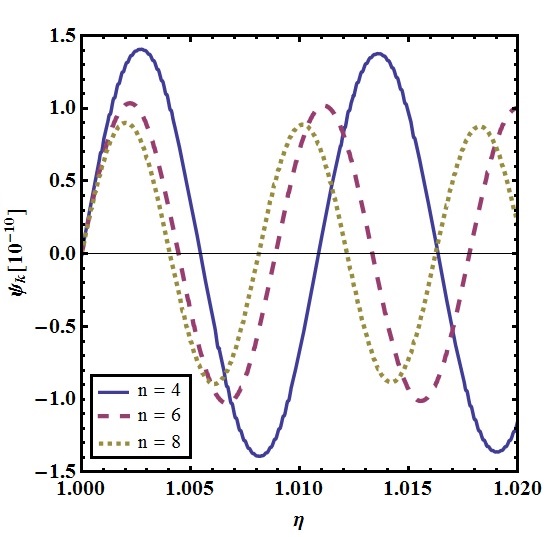}
\caption{\footnotesize{In this figure, we compute the metric perturbation, $\Psi_k$, by the non averaged system  (\ref{Eij}-\ref{Eii}) for a scalar theory with various potentials: $V = \lambda \vert\phi\vert^n/n \;;$ $\;n=\{ 4,\;6,\;8 \}$. We work in the same units of Fig. 1. The parameters used for each case were respectively: $\lambda =\{10^{32},\;10^{52},\;10^{64} \}$, $k = 1000$; and we consider the initial conditions: $\delta\phi_k(1)=10^{-7}$, $\dot{\delta\phi}_k(1)= - 3 \cdot 10^{-7}$ and $a(1) = 1$. This numerical result is well described by (\ref{ceffProp}) with $c_{\text{eff}}^2 = \{1/3,\;1/2,\;3/5\}$ respectively.}}
\label{Eceff4}
\end{figure}
\begin{figure}[h]
\includegraphics[width= 0.85\columnwidth]{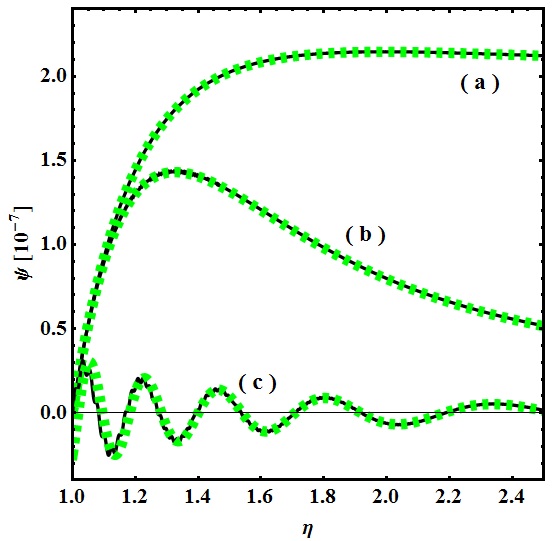}
\caption{\footnotesize{In this figure, we compare the metric perturbation, $\Psi_k$, computed by the non averaged system (\ref{Eij}-\ref{Eii}) (continuous line) and by the effective equation (\ref{ceffProp}) (dashed line) for a massive scalar field. (a),(b) and (c) label the cases $m = \{3\cdot 10^3, 10^3, 100\}$,  respectively, with $k = 100$. We work in the same units of Fig. 1. The initial conditions for the exact system are $\delta\phi_k(1)=10^{-6}$, $\dot{\delta\phi}_k(1)=-6 \cdot 10^{-6}$ and $a(1)=1$. The initial conditions for the effective equation has been tuned
properly. We observe good concordance between both, moreover it can be seen that the shape of (b) and (c) comes from the correction given
by Eq. (\ref{Csk}).}}
\label{Eceff2}
\end{figure}

\section{Perturbation of power law potential theories}

In general for a power-law potential, the background field equation is non linear  and we cannot obtain the explicit adiabatic expansion of the solutions as in the quadratic
case. We can still define a typical frequency of the background oscillations $\omega_{\text{eff}}$ and compare it with the other two scales in the 
problem $\mathcal{H}$ and $k$. Moreover, in this case, the pressure is not negligible in comparison with the energy density. This fact makes easy to compute the effective sound speed $c_{\text{eff}}$, which is the quantity that controls the evolution of sub-Hubble perturbations.

\subsection*{Effective sound speed}

Combining (\ref{PEins00}) and (\ref{PEinsii}), we get a single equation
for the metric perturbation:
\begin{eqnarray}
&&\ddot{\Psi}_k + 3 \mathcal{H} \left( 1 + c^2_{\text{eff}} \right) \dot{\Psi}_k + c^2_{\text{eff}} k^2 \Psi_k\nonumber
\\
&& \;\;\;\;\;\;+ \left[ 2 \dot{\mathcal{H}} + \left( 1 + 3 c^2_{\text{eff}} \right)\mathcal{H}^2 \right] \Psi_k =0\;.
\label{ceffProp}
\end{eqnarray}
 If $\phi$ dominates the background energy density, (\ref{ceffProp}) can be rewritten as
\begin{eqnarray}
\ddot{\Psi}_k &+& 3 \mathcal{H} \left( 1 + c^2_{\text{eff}} \right) \dot{\Psi}_k
\\
&+& \left(c^2_{\text{eff}} k^2 +3 \left(c^2_{\text{eff}}-\omega\right)\mathcal{H}^2 \right) \Psi_k \;, \nonumber
\end{eqnarray}
where here, $\cal H$ and $\omega$ correspond to the average background evolution.

In order to compute the effective sound speed, let us consider the following average in a period $T$, which verifies ${\cal H}^{-1} \gg T \gg \omega_{\text{eff}}^{-1} $:
\begin{eqnarray}
&& \left \langle \partial_0 \left( \left( \dot{\phi} + \dot{\delta\phi}_k \right) (\phi + \delta\phi_k)\right) \right\rangle \equiv \nonumber
\\
\nonumber
\\
&& \frac{\left( \dot{\phi} + \dot{\delta\phi}_k \right) (\phi + \delta\phi_k)\vert_{t' = t+ T}- \left( \dot{\phi} + \dot{\delta\phi}_k \right) (\phi + \delta\phi_k)\vert_{t' = t}}{T} \nonumber
\\
\nonumber
\\
&& = \left\langle \left( \dot{\phi} + \dot{\delta\phi}_k \right)^2 + \left( \ddot{\phi} + \ddot{\delta\phi}_k \right)\left( \phi + \delta\phi_k \right)\right\rangle\;.
\end{eqnarray}
If the field evolution is periodic or bounded, averaging during long enough  periods, the left-hand term of the last equation will be negligible in comparison with the average of $\left( \dot{\phi} + \dot{\delta\phi}_k \right)^2$.

In principle, except for a quadratic potential, $\delta\phi$ will have some growing modes. Those modes make $ \left \langle\partial_0 \left( \dot{\phi}\delta\phi + \phi \dot{\delta\phi}\right)\right\rangle$ not to vanish, but as long as it oscillates around zero, the following discussion holds at leading order.

Focusing on the first order of perturbations and introducing the equation of motion (\ref{Eqdphi}):
\begin{eqnarray}
&&\langle \dot{\delta \phi_k} \dot{\phi} \rangle = \left\langle - 2 \dot{\Psi}_k \dot{\phi} \phi + \frac{k^2}{2} \delta\phi_k \phi + \frac{a^2}{2} V'(\phi) \delta \phi_k \right \rangle \nonumber
\\
&+& \left \langle \frac{a^2}{2} V''(\phi) \; \phi \delta \phi_k + a^2 V'(\phi) \Psi_k \phi \right\rangle+\Od(\epsilon)
\;;\label{Pvirial}
\end{eqnarray}
where we have neglected the scale factor derivatives as we are considering time intervals much smaller than the inverse of the expansion rate.

By using (\ref{Pvirial}), we obtain the following expression for the effective speed of sound
\begin{widetext}
\begin{eqnarray}
c_{\text{eff}}^2 &=& \frac{\left\langle \frac{k^2}{a^2} \delta \phi_k\; \phi - V'(\phi) \delta\phi_k + V''(\phi) \phi \delta\phi_k  - \frac{2}{a^2} \dot{\Psi}_k \partial_0 (\phi^2)  - 2 \Psi_k \left(  \frac{\dot{\phi}^2}{a^2}   - V'(\phi) \phi\right) \right\rangle}{\left\langle \frac{k^2}{a^2} \delta \phi_k\; \phi + 3 V'(\phi) \delta\phi_k + V''(\phi) \phi \delta\phi_k - \frac{2}{a^2} \dot{\Psi}_k \partial_0 (\phi^2)  - 2 \Psi_k \left(  \frac{\dot{\phi}^2}{a^2} - V'(\phi) \phi\right) \right\rangle}\nonumber
\\
 &\simeq& \frac{\left\langle \frac{k^2}{a^2} \delta \phi_k\; \phi - V'(\phi) \delta\phi_k + V''(\phi) \phi \delta\phi_k\right\rangle}{\left\langle \frac{k^2}{a^2} \delta \phi_k\; \phi +3 V'(\phi) \delta\phi_k + V''(\phi) \phi \delta\phi_k\right\rangle}+ \mathcal{O}\left(\epsilon\right)\;.\;\;\;\;\; \label{Cseff}
\end{eqnarray}
\end{widetext}
In the second equation the terms proportional to $\Psi_k$ are averaged out as their fast oscillating part can be expressed as a total derivative or coincide with the background generalization of the virial theorem (\ref{virial}).


If a power-law potential, $V(\phi) = \lambda \vert\phi\vert^n/n$, is considered:

\begin{itemize}
\item When $ \omega_{\text{eff}} \gg k$, i.e. the frequency of $\phi$ is almost equal to the perturbation frequency, the behaviour is similar to a perfect fluid with
constant equation of state (see Fig. \ref{FigsubH} for the case $n=2$ and  Fig. \ref{Eceff4} for $n= \{ 4,\;6,\;8 \} $), i.e. 
up
to $\Od(\epsilon)$:
\begin{eqnarray}
c_{\text{eff}}^2 = \frac{n - 2}{n + 2} = \omega \equiv \frac{\langle p \rangle }{\langle \rho \rangle }\;.
\end{eqnarray}
Notice that for $n<2$, there is an instability  in agreement with \cite{Kamionkowski}. In this case $V''(\phi)$ is not well defined when $\phi = 0$.

\item For massive scalar fields $n = 2$,
\begin{eqnarray}
\langle \delta p_k\rangle &=& \left \langle \frac{k^2}{a^2} \delta\phi_k \; \phi \right\rangle +\Od(\epsilon)\;,
\\
\langle \delta\rho_k \rangle &=& \left \langle \frac{k^2}{a^2} \delta\phi_k \; \phi + 4 m^2 \phi \; \delta\phi_k \right\rangle+\Od(\epsilon) \;.
\end{eqnarray}
Reproducing the result obtained in \cite{Khlopov-1985,Ratra-1991,Hwang}:
\begin{eqnarray}
c_{\text{eff}}^2 &=& \frac{k^2}{k^2 + 4 m^2 a^2}+\Od(\epsilon)\simeq\frac{k^2}{4 m^2 a^2}+\Od(\epsilon)\;. \label{Csk}
\end{eqnarray}
Although this expression seems to be valid for all $k$, when $k$ is comparable with $m a$ fails since the averaging time interval has the same order of the $\delta \rho$ oscillation period. See curve (c) in Fig. \ref{Eceff2} at early times. The high $k$ limit will be discussed in section VIII.

The corresponding comoving Jeans length $\lambda_J=2\pi/k_J$, which satisfies
$c_{\text{eff}}^2(k_J)k_J^2=4\pi G\langle\rho\rangle a^2$ reads:
\begin{eqnarray}
k_J^2 = \sqrt{8\pi G}\phi_c m^2a^2\;. \label{jeans}
\end{eqnarray}
\item Because of the vanishing of the potential terms in the numerator
of (\ref{Cseff}) in the harmonic case, we can also consider anharmonic corrections, $V(\phi) = m^2 \phi^2/2+ \lambda \phi^l /l$ in 
a simple way.  
 Due to the averaging process  the sine mode of $\delta\phi_k$ gives 
 a negligible contribution:
\begin{eqnarray}
&& \left \langle \phi^{l-1} \sin\left(\int m a\; d\eta\right) \right\rangle = \phi_c^{l-1}(\eta)
\\
&& \cdot \left \langle \partial_0 \left(\frac{\cos\left(\int m a\; d\eta\right)^l}{l}\right) \right\rangle \sim \mathcal{O}\left(\epsilon\phi_c^{l-1}(\eta)\right)\;.\nonumber
\end{eqnarray}
The same situation occurs with the term $k^2\phi \delta\phi_k$. Therefore, only the cosine mode contributes and we will be able to extract the common factor $\phi_0 \delta\phi_c$ both in the 
numerator and the denominator and simplify the expression. We will only consider even powers $l = 2 p$ since odd exponents  make the sinusoidal resulting function to oscillate around zero and are suppressed by the average. Finally,
\begin{eqnarray}
c_{\text{eff}}^2 = \frac{k^2}{4 m^2 a^2} + \frac{\left(p -1\right)}{2^{2 p}} \left(\begin{array}{c}
2p
\\
p
\end{array}\right) \frac{\lambda \phi_c^{2p-2}}{m^2 a^{3(p-1)}}
+\Od(\epsilon)\;.
\end{eqnarray}
\end{itemize}
For example, for the case $p = 2$, we recover the sound speed obtained in \cite{Kamionkowski}.

\section{Comparing with the non-averaged solutions}
In order to check the validity of the results 
obtained above from the averaged equations, in
this section 
we will compare them with  exact results 
in certain particular limits.

Let us then consider the Einstein equations
with the exact (non-averaged) energy-momentum tensor.
At the background level we have:
\begin{eqnarray}
\mathcal{H}^2 = \frac{8\pi Ga^2}{3}\rho = \frac{8\pi Ga^2}{3} \left( \frac{\dot{\phi}^2}{2 a^2} + V(\phi)\right)\;; \label{ExEins00}
\\
2 \dot{\mathcal{H}} + \mathcal{H}^2 =  - 8\pi G a^2 p =- 8\pi G a^2\left( \frac{\dot{\phi}^2}{2 a^2} - V(\phi) \right)\; \label{ExEinsii}
\end{eqnarray}
whereas to first order in  perturbations they read,
\begin{eqnarray}
&&  \Psi_k - \Phi_k = 0\;, \label{Eij}
\\
\nonumber
\\
&& - 6 \mathcal{H}^2 \Psi_k - 6 \mathcal{H}  \dot{\Psi}_k - 2 k^2 \Psi_k =8\pi G a^2 \delta \rho_k \;, \label{E00}
\\
\nonumber
\\
&& 2 \mathcal{H} \Psi_k + 2 \dot{\Psi}_k =-i8\pi G a^2  k_i\frac{\delta T^0_{\;i}\vert_k}{k^2}\;, \label{E0i}
\\
\nonumber
\\
&& \ddot{\Psi}_k + 3 \mathcal{H} \dot{\Psi}_k + \left( \mathcal{H}^2 + 2 \dot{\mathcal{H}}\right) \Psi_k= 4\pi G a^2 \delta p_k \;, \label{Eii}
\end{eqnarray}
where the expression for the perturbed energy-momentum tensor components are 
given in (\ref{dE00}), (\ref{dEii}) and (\ref{dE0i}). 
Notice that in this section all the geometric quantities are not
averaged so that in order to compare with the result of the 
previous section we will explicitly take the corresponding average 
of the obtained results.

\subsection{Super-Hubble analytic approach}

In the super-Hubble limit  $k \ll \mathcal{H}$, we can neglect the terms proportional to $k$ in (\ref{E00}),
and introducing (\ref{E0i}) in (\ref{E00}), we obtain
\begin{eqnarray}
- 3 \mathcal{H} \dot{\phi} \delta\phi_k  = \dot{\phi}\dot{\delta\phi_k} - \Psi_k \dot{\phi}^2+ V'(\phi) \delta\phi_k\;.
\end{eqnarray}

\begin{figure*}[t]
\includegraphics[width=\textwidth]{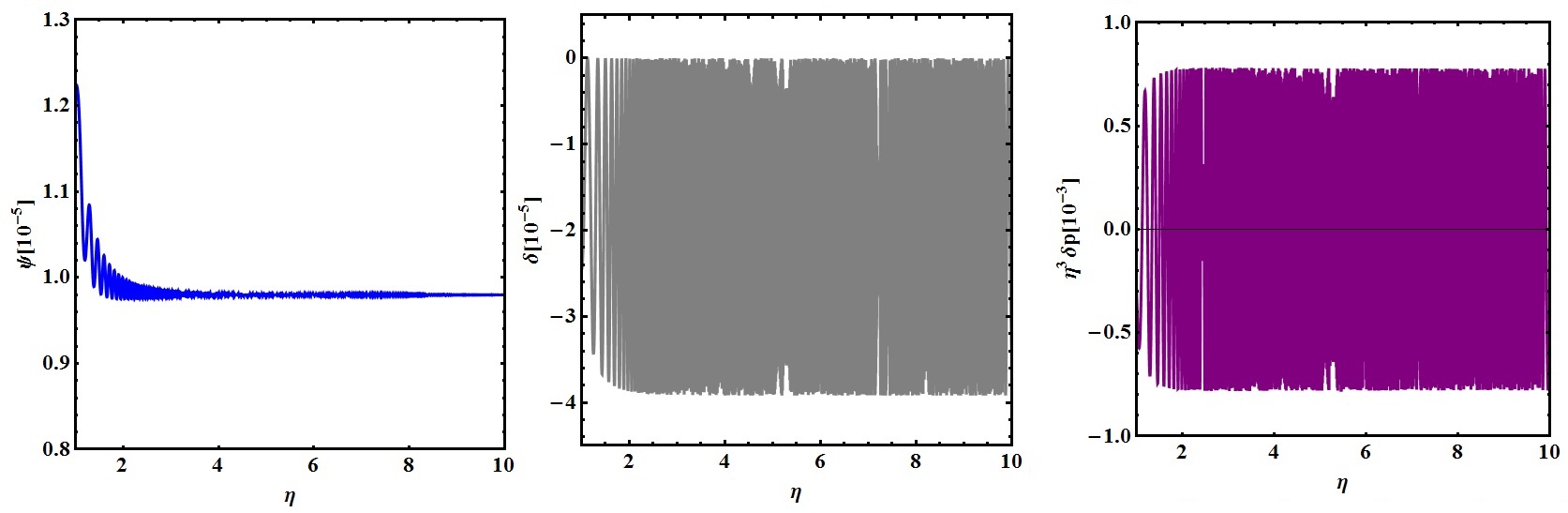}
\caption{\footnotesize{Super-Hubble mode of a massive scalar field. The plots are the numerical solution to the non averaged equations (\ref{Eij}$-$\ref{E0i}), with $V(\phi)= m^2 \phi^2/2$. We work in the same units of Fig. 1. In order to calculate a representative mode we have set $m = 10$, $k=0$, $\delta\phi_k(\eta=1)= 10^{-5}$, $\dot{\delta\phi}_k(\eta=1) = 0$, $a(\eta=1)= 1$. It can be seen that the scalar perturbation of the metric (blue) tends to a constant. The perturbation of the energy density (gray) decays as $\eta^{-6} \sim a^{-3}$, consequently the average contrast of density is constant as expected for the perturbations of a dust-like perfect fluid during the super-Hubble limit. Finally, the pressure perturbation (purple) oscillates around zero.}}
\label{FigsuperH}
\end{figure*}

It is interesting to note that any time the oscillating $\dot{\phi}$ is zero, the perturbation $\delta \phi_k$ must be zero too.
Therefore, it is natural to assume the ansatz $\delta\phi_k = f_k(\eta) \dot{\phi}$. When it is substituted in the previous equation
and by using Eq. (\ref{Eqphi}), we can write the metric  perturbation as:
\begin{eqnarray}
\Psi_k = \dot{f}_k(\eta) + \mathcal{H} f_k(\eta)\;.
\label{PsiSH}
\end{eqnarray}
If we take into account (\ref{E0i}), we can obtain the following equation for $f_k(\eta)$:
\begin{eqnarray}
\ddot{f}_k(\eta) + 2 \mathcal{H} \dot{f}_k(\eta) + \left(\mathcal{H}^2 + \dot{\mathcal{H}} \right) f_k(\eta) = 4\pi G f_k(\eta) \dot{\phi}^2 .
\end{eqnarray}
By combining equations (\ref{ExEins00}) and (\ref{ExEinsii}), we can integrate the last equation obtaining:
\begin{eqnarray}
\dot{f}_k(\eta) = - 2 \mathcal{H} f_k(\eta) + \tilde{c}^0_k\;,
\end{eqnarray}
where $\tilde{c}^0_k$ is an integration constant. Solving this equation we obtain
\begin{eqnarray}
f_k(\eta) = \frac{c^1_k \eta_0}{a} + \frac{c^0_k}{a^2} \int^\eta_{\eta_0} a^2(\eta') d\eta'\;,\label{fex}
\end{eqnarray}
where $c^0_k$ and $c^1_k$ are integration constants. Substituting in (\ref{PsiSH})
\begin{eqnarray}
 \Psi_k= c^0_k \left(1-\frac{\mathcal{H}}{a^2} \int^\eta_{\eta_0} a^2(\eta') d\eta'\right)- \frac{c^1_k \eta_0 \mathcal{H}}{ a^2}\;.
\label{Psiex}
\end{eqnarray}
As it can be seen in Fig. \ref{FigsuperH} for the case of a quadratic potential,
$\Psi_k$ does not oscillate in an important way. On the other hand, computing the energy density contrast:
\begin{eqnarray}
\delta_k \equiv \frac{\delta \rho_k}{\rho} &=& - 3 \mathcal{H} f(\eta) \frac{\dot{\phi}^2}{\rho} 
= - 3 \mathcal{H} f(\eta)\frac{\rho + p}{\rho}\label{deltaex}\;,
\end{eqnarray}  
Notice that since $\dot{\phi}^2$ oscillates around a non-vanishing
value, the same behaviour is expected for $\delta_k$. This agrees  with  the numerical result in Fig. \ref{FigsuperH}.

Finally, it can be seen that averaging the previous expression, 
we find a perfect agreement with the results in previous section. Thus, substituting (\ref{a}) in (\ref{Psiex}) and (\ref{fex}), and averaging in (\ref{deltaex}) we find that at late times:
\begin{eqnarray}
\langle \delta_k\rangle = -\frac{6 (1 + \omega)}{3 \omega + 1} c^0_k= -2 \langle \Psi_k\rangle +\Od(\epsilon)\;, \label{psiSuperH}
\end{eqnarray}
where the $\Od(\epsilon)$ error comes from the average in $\dot\phi^2$.  
çThis result agrees with the standard expression for super-Hubble modes in perfect fluid cosmologies with constant equation of state. Notice that (\ref{psiSuperH}) is in good concordance with the numerical solution shown in Fig. \ref{FigsuperH}.

\subsection{Sub-Hubble analytic approach}

In this section, we will consider a different approach to check the results of previous section. This approach is valid for sub-Hubble modes of power law potential theories with $ n \neq 2$.  Fortunately, as we shown, the $n = 2$ case is well understood.

Using (\ref{Eqphi}) and (\ref{Eqdphi}), we can write the perturbation of the energy density and pressure as,
\begin{eqnarray}
\delta \rho_k &=& \frac{1}{2 a^2} \left( \dot{\Sigma}_k- \frac{n + 2}{n - 2}  \dot{\Delta}_k - \frac{4}{n - 2} k^2 \phi \delta \phi_k \right) \nonumber
\\
&-& \left( \frac{\dot{\phi}^2}{a^2} + \frac{4 n}{n-2} \lambda \phi^n \right) \Psi_k  + \frac{4 \partial_0 \left(\phi^2\right)}{(n -2)a^2}\dot{\Psi}_k   \nonumber
\\
&-& \frac{\mathcal{H}}{(n-2)a^2} \left( 4 \phi \dot{\delta \phi}_k - 2 n \delta\phi_k \dot{\phi}\right)\;, \label{deltarho}
\\
\delta p_k &=& \frac{1}{2 a^2} \left( \dot{\Sigma}_k- \dot{\Delta}_k \right) - \frac{\dot{\phi}^2}{a^2} \Psi_k + 2 \mathcal{H} \frac{\dot{\phi} \delta\phi_k}{a^2}\;,\label{deltap}
\end{eqnarray}
where $\dot{\Sigma}_k \equiv \partial_0^2 \left( \phi \; \delta\phi_k \right)$, $\dot{\Delta}_k \equiv  \partial_0 \left( \phi \; \dot{\delta\phi}_k - \dot{\phi}\; \delta\phi_k\right)$.

Multiplying (\ref{E00}) by the expected effective sound speed $c_{\text{eff}}^2 = (n - 2)/(n + 2)$ and subtracting it from (\ref{Eii}):
\begin{eqnarray}
\ddot{\Psi}_k &+& 3 \mathcal{H} \left( 1 + c^2_{\text{eff}} \right) \dot{\Psi}_k + \left(c_{\text{eff}}^2 k^2 + 2 \dot{\mathcal{H}} \right.
\\
&+&\left. \left(1 + 3 c_{\text{eff}}^2\right)\mathcal{H}^2\right) \Psi_k = 4 \pi G a^2 \left( \delta p_k - c_{\text{eff}}^2 \delta \rho_k \right)\;.\nonumber
\end{eqnarray}

By using (\ref{deltarho}) and (\ref{deltap}),
\begin{eqnarray}
\label{eqjo}
\ddot{\Psi}_k &+& 3 \mathcal{H} \left( 1 + c^2_{\text{eff}} + \frac{16 \pi G}{3 \left(n + 2\right)} \frac{\partial_0\left(\phi^2\right)}{\mathcal{H}}\right) \dot{\Psi}_k
\\
&+&  c_{\text{eff}}^2 k^2 \Psi_k = \frac{8 \pi G}{n + 2} \left( \partial_0^2 +\mathcal{H} \partial_0 + k^2\right) \phi \;\delta\phi_k\;.\nonumber
\end{eqnarray}
Notice that the order of the $\phi$ amplitude can be estimated through (\ref{Eins00}),
\begin{eqnarray}
\phi_c \sim \mathcal{O}\left(\frac{\mathcal{H}}{\sqrt{8 \pi G} \omega_{\text{eff}}}\right)\;,
\end{eqnarray}
thus,
\begin{eqnarray}
\frac{16 \pi G}{3 \left(n + 2\right)} \frac{\partial_0\left(\phi^2\right)}{\mathcal{H}} \sim \mathcal{O}\left(\frac{\mathcal{H}}{\omega_{\text{eff}}}\right) \ll 1\;.
\end{eqnarray}
So Eq. (\ref{eqjo}) can be approximated by
\begin{eqnarray}
\ddot{\Psi}_k &+& 3 \mathcal{H} \left( 1 + c^2_{\text{eff}}\right) \dot{\Psi}_k +c_{\text{eff}}^2 k^2 \Psi_k \label{exacddPsi}
\\
&=& \frac{8 \pi G}{n + 2} \left( \partial_0^2 +\mathcal{H} \partial_0 + k^2\right) \phi \;\delta\phi_k\;.\nonumber
\end{eqnarray}

Moreover, in time intervals $\Delta \eta \ll \mathcal{H}^{-1} $, we can write approximately
\begin{eqnarray}
\ddot{\Psi}_k + c_{\text{eff}}^2 k^2 \Psi_k &=& \frac{8 \pi G}{n + 2} \left(\dot{\Sigma} + k^2 \phi \delta \phi_k \right) \label{EcMinPsi}
\\
&=& \frac{8 \pi G}{n + 2} \left(\partial_0^2 + k^2 \right) \phi \delta \phi_k  \;.\nonumber
\end{eqnarray}

The solution of this equation is decomposed in a solution for the homogeneous equation,
\begin{eqnarray}
\Psi_{k, \text{hom}} (\eta) = \Psi_{0k} \; \cos \left( c_{\text{eff}} k \eta + \Delta_{0k} \right)\;,
\end{eqnarray}
and a particular solution for the inhomogeneous one. In order to obtain it let us take the Fourier transform of (\ref{EcMinPsi}),
\begin{eqnarray}
{\hat\Psi}_{k, \text{part}} (\omega) &=& - 8 \pi G \frac{\omega^2 - k^2}{\omega^2 - c_{\text{eff}}^2 k^2} \, \hat{\phi}*\hat{\delta \phi}_k \;.
\end{eqnarray}
where $*$ denotes the convolution operation.

In the limit $\omega_{\text{eff}} \gg k$, $\phi \delta\phi_k$ oscillates with $\omega_{\text{eff}}$ and, thus,
\begin{eqnarray}
{\hat\Psi}_{k, \text{part}} (\omega) = - 8 \pi G\; \hat{\phi}*\hat{\delta \phi}_k \left(1 + \mathcal{O}\left(\frac{k^2}{\omega^2}\right)\right)\;.
\end{eqnarray}
By taking the inverse Fourier transform, we reach the general solution,
\begin{eqnarray}
\Psi_k (\eta) &\simeq& \Psi_{0k}  \cos \left( \frac{2 c_{\text{eff}}\, k}{\omega_{\text{eff}}} z + \Delta_{0k} \right) -  8 \pi G \phi \delta \phi_k\;,
\end{eqnarray}
where $z = \omega_{\text{eff}}\, \eta/2$.

As it can be seen, $\Psi_{k, \text{part}} \sim \mathcal{O} (8 \pi G \phi \delta \phi_k)$, however for sub-Hubble modes,

\begin{eqnarray}
\Psi_k \simeq \frac{8 \pi G a^2 \delta\rho_k}{k^2}\sim \mathcal{O} \left(8 \pi G \frac{\omega_{\text{eff}}^2}{k^2} \phi \delta\phi_k\right)\;, \label{psiorder}
\end{eqnarray}
and, consequently, the particular solution can be neglected. In addition, the source on (\ref{exacddPsi}) can also be neglected, thus obtaining the same approximated equation (\ref{ceffProp}) of the effective model.

A more rigorous analysis can be done thanks to Floquet's theorem. Within time intervals $\Delta \eta \ll \mathcal{H}$, (\ref{Eqdphi}) takes the form,
\begin{eqnarray}
\partial_z^2 \delta \phi_k + \left(Q(z) +\epsilon_k^2 \right) \delta \phi_k =0\;.\label{EcpertFlo}
\end{eqnarray}
with $Q(z)\equiv 4 (n-1)\lambda \vert \phi \vert^{n-2}/\omega_{\text{eff}}^2$ and $\epsilon_k \equiv 2 k / \omega_{\text{eff}}$. Notice that as $\phi$ is periodic, so is $Q(z)$ and the Floquet's theorem gives us the general form of
the solution of (\ref{EcpertFlo}).

 Following \cite{Magnus}, let us define $\delta\phi^{(1)}_k$ and $\delta\phi^{(2)}_k$ as the normalized solutions corresponding to the initial conditions: $\delta\phi^{(1)}_k(0) =1$, $\dot{\delta\phi}^{(1)}_k(0)=0$ and $\delta\phi^{(2)}_k(0) =0$, $\dot{\delta\phi}^{(2)}_k (0)=1$. The corresponding characteristic equation reads
\begin{eqnarray}
\tau^2-\left[ \delta\phi^{(1)}_k(\pi)+\dot{\delta\phi}^{(2)}_k (\pi)\right] \tau + 1 = 0\;,
\end{eqnarray}
whose associated roots can be written as
\begin{eqnarray}
\tau_{1k}=e^{i \alpha_k \pi},\;\tau_{2k}=e^{-i \alpha_k \pi}\;,
\end{eqnarray}
and the  Floquet's theorem implies:
\begin{enumerate}
  \setcounter{enumi}{0}
  \item If $\tau_{1k} \neq \tau_{2k}$, (\ref{EcpertFlo}) has two linearly independent solutions
  \begin{eqnarray}
 \delta\phi^{(1)}_k = e^{i \alpha_k z} \xi_{1k}(z)\;,\;\delta \delta\phi^{(2)}_k = e^{- i \alpha_k z} \xi_{2k}(z)\;,
  \end{eqnarray}
  where $\xi_{1k}$ and $\xi_{2k}$ are $\pi$-periodic functions.

   \item If $\tau_{1k} = \tau_{2k}$, (\ref{EcpertFlo}) has a periodic solution with period $\pi$ ($\tau_{1k}=\tau_{2k}=1$) or $2 \pi$ ($\tau_{1k}=\tau_{2k}=-1$). Denoting by $\xi(z)$ this one and by $\delta\phi^{(2)}_k(z)$ another linearly independent solution:
   \begin{eqnarray}
   \delta\phi^{(2)}_k(z + \pi) = \tau_{1k} \delta\phi^{(2)}_k(z)+ c \; \xi_k(z)\;, \label{Floquet2}
   \end{eqnarray}
   where $c$ is a constant.
\end{enumerate}

\begin{figure}[t]
\includegraphics[width=\columnwidth]{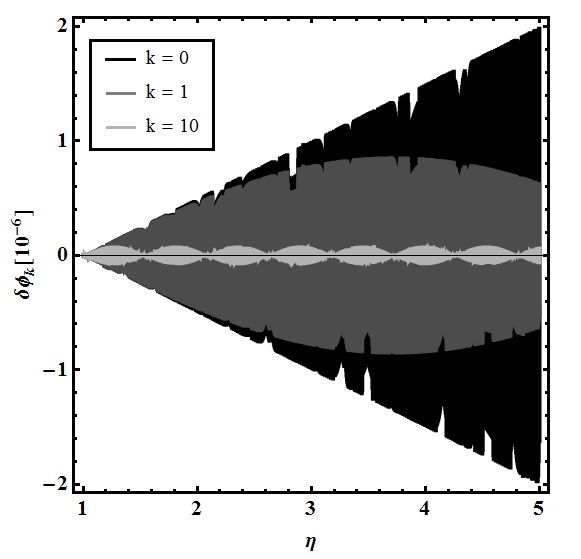}
\caption{\footnotesize{This plot shows the numerical solutions of (\ref{PFloquet4}) for different k values. We work in the same units of Fig. 1. The initial conditions are $\delta\phi_k (0) = 0$ and  $\dot{\delta\phi}_k (0) = - 10^{-6}$, for $\lambda = 10^6$. The solution with $k = 0$ (black) shows the growing mode of (\ref{dphi4k0}), which corresponds to the second case of Floquet's theorem. As it can be seen for the cases with $k = 1$ (gray) and $k = 10$ (lighter gray), in general the characteristic equation roots will be different and the first case of Floquet's theorem applies.}}
\label{FigFloquet}
\end{figure}

As an example, let us consider the quartic potential case $V(\phi) = \lambda \phi^4 /4$,
\begin{eqnarray}
\phi &=& \phi_0  \text{sn}[z,i]\;;\;z = \sqrt{\frac{\lambda}{2}}\phi_0 \eta + \Delta\;;
\\
\ddot{\delta\phi}_k &+& \left( 6\; \text{sn}[z,i]^2 + \epsilon_k^2 \right) \delta\phi_k =0\;;\;\; \epsilon_k \equiv \frac{2 k^2}{\lambda \phi_0^2}\;,\label{PFloquet4}
\end{eqnarray}
with $\phi_0$ and $\Delta$ integration constants and $\text{sn}[z,i]$ the corresponding Jacobi Elliptic function. If $k =0$,
\begin{eqnarray}
\delta \phi_{k=0} = \delta\phi_{10} \frac{d}{dz}\left( \text{sn}[z,i] \right)+\delta\phi_{20} \frac{d}{dz}\left( z \cdot \text{sn}[z,i] \right)\;. \label{dphi4k0}
\end{eqnarray}
Notice that, as Jacobi Ellliptic functions are periodic with period $4 K$,
\begin{eqnarray}
K \equiv \int\limits_{0}^{\frac{\pi}{2}} \frac{d \theta}{\sqrt{1+\sin^2(\theta)}}
\end{eqnarray}
making the change $z =4 K x /\pi $, we can see that this solution corresponds to the $\tau_{1k} = \tau_{2k} = 1$ case of the Floquet's theorem:
\begin{eqnarray}
\delta \phi_k (z + \pi) &=& \delta\phi_{10} \frac{\pi}{4 K} \frac{d}{dx}\left( \text{sn} \left[ \frac{4 K x}{\pi},i \right] \right) \nonumber
\\
&+& \delta\phi_{20} \left( \frac{\pi}{4 K} \frac{d}{dx}\left( \frac{4 K x}{\pi} \text{sn}\left[\frac{4 K x}{\pi},i \right] \right)\right. \nonumber
\\
&+& \left. 4 K \frac{d}{dx}\left( \text{sn}\left[\frac{4 K x}{\pi},i\right] \right) \right)\;,
\end{eqnarray}
by comparing with (\ref{Floquet2}). However, in general, when $k \neq 0$, $\tau_{1k} \neq \tau_{2k}$ and it can be seen that for modes with $\omega_{eff} \gg k$, $\delta \phi$ is an oscillating function modulated by a periodic function of frequency $\mathcal{O} \left( k /\sqrt{\lambda \phi_0^2} \right)$. In Fig. \ref{FigFloquet} an explicit numerical calaculation shows how the number of nodes is multiplied by a factor of 10 when $k$ is increased from $k = 1$ to $k = 10$.

For a general power law potential, if $\epsilon_k = k = 0$, the analogous solution to (\ref{dphi4k0}) can be written as
\begin{eqnarray}
\delta\phi_{k=0} = \delta\phi_{10}\dot{\phi}+\delta\phi_{20}\left( \frac{2}{n-2} \phi + t \dot{\phi}\right)\;,
\end{eqnarray}
which also corresponds to the second case of Floquet's theorem. For $k\neq 0$, $\delta \phi_k$ behaves analogously to the $n = 4$ case and, thus, we  will assume that $\delta\phi_k$ has the form:
\begin{eqnarray}
\delta \phi_k (z) =  e^{i \alpha_k z} \xi_{1k}(z) +  e^{- i \alpha_k z} \xi_{2k}(z)\;.
\end{eqnarray}

As $\delta\phi_k \in \mathbb{R}$, solutions can be divided in \cite{Arscott}:

\begin{enumerate}
\item Stable solutions with $\text{Re}(\alpha_k) = 0$:
\begin{eqnarray}
\delta \phi_k (z) =  e^{i \alpha_k z} \xi_{1k}^*(z) +  e^{- i \alpha_k z} \xi_{1 k}^*(z)\;.
\end{eqnarray}

\item Unstable solutions with $\text{Re}(\alpha_k)=\mu$ and $\text{Im}(\alpha_k)= l \in \mathbb{Z}$:
\begin{eqnarray}
\delta \phi_k (z) = e^{\mu_k z} e^{i l z} \xi_{1k}(z) +  e^{-\mu_k z} e^{-i l z} \xi_{2k}(z)\;.
\end{eqnarray}
Notice that $e^{i l z} \xi_{1k}(z)$ and $e^{-i l z} \xi_{2k}(z)$ are real $\pi$-periodic functions.
\end{enumerate}

We can synthesise both cases in a single expression:
\begin{eqnarray}
\delta \phi_k =    e^{i \alpha_k z} \tilde{\xi}_{1k}(z) +  e^{-i \alpha_k z} \tilde{\xi}_{2k}(z)\;,
\end{eqnarray}
where $\tilde{\xi}_{1k}(z)$ and $\tilde{\xi}_{2k}(z)$ are  $\pi$-periodic functions. We consider $\text{Im}(\alpha_k) = 0$, and $\tilde{\xi}_{1k} = \tilde{\xi}_{2k}^*$ for stable modes, which are those in which we are interested.

Due to the periodic properties of $\phi$ and $\delta \phi_k$, their product can be expanded in Fourier series as,
\begin{eqnarray}
8 \pi G \phi \delta\phi_k &=&  e^{i \alpha_k z} \sum\limits^{+\infty}_{m = - \infty} b_{mk} e^{i 2 m z}
\\
&+&  e^{-i \alpha_k z} \sum\limits^{+\infty}_{m = - \infty} b_{mk}^* e^{-i 2 m z}\;,  \nonumber
\end{eqnarray}
 From this series, we can obtain a simple expression for $\hat{\phi} * \hat{\delta \phi}_k$ ,
\begin{eqnarray}
\hat{\Psi}_{k, \text{part}} (\omega) &=& - \frac{\omega^2 - \frac{4 k^2}{\omega_{\text{eff}}^2}}{\omega^2 - \frac{4 k^2}{\omega_{\text{eff}}^2} c_{\text{eff}}^2} 
\\
&\times& \left(  \sum\limits^{+\infty}_{m = - \infty} b_{mk}  \delta \left(\omega - i(\alpha_k + 2 m) \right) \;\;\;\;\;\right.\nonumber
\\
&+& \left.  \sum\limits^{+\infty}_{m = - \infty} b_{mk}^*  \delta \left(\omega + i (\alpha_k + 2 m) \right) \right)\;,\;\nonumber
\end{eqnarray}
We can obtain the general solution of (\ref{EcMinPsi}),
\begin{eqnarray}
&& \Psi_k = \Psi_{0k}  \cos\left( \frac{2 c_{\text{eff}}\, k}{\omega_{\text{eff}}} z +\Delta_{0k} \right)\label{analyticPsi}
\\
&-&  \sum\limits^{+\infty}_{m = - \infty} c_{mk} b_{mk} e^{i( 2m + \alpha_k) z} - \sum\limits^{+\infty}_{m = - \infty} c_{mk} b_{mk}^* e^{-i( 2m + \alpha_k) z}\;,\nonumber
\end{eqnarray}
with
\begin{eqnarray}
c_{mk}&=& \frac{(\alpha_k + 2 m)^2 +\frac{4 k^2}{\omega_{\text{eff}}^2}}{(\alpha_k + 2 m)^2 + \frac{4 k^2}{\omega_{\text{eff}}^2} c_{\text{eff}}^2},
\end{eqnarray}
Notice that in the limit in which we are interested ($\omega_{\text{eff}}\gg k$), $c_{mk}  \simeq 1$. Therefore the particular solution is at leading order equal to $\phi \delta\phi_k$ and is negligible in comparison with the homogeneous one, in agreement with our preliminary analysis (\ref{psiorder}).

From the general solution (\ref{analyticPsi}), we can obtain an 
expression for the effective sound speed up to $\Od(\epsilon)$:
 \begin{widetext}
 \begin{eqnarray}
 c_{\text{eff}}^2 \equiv  \frac{\langle \delta p_k \rangle}{\langle \delta \rho_k \rangle}
  \simeq \frac{ \left\langle \frac{\omega_{\text{eff}}^2}{4} \partial_z^2 \Psi_k (z) \right \rangle}{\left \langle - k^2 \Psi_k (z) \right \rangle}
  \simeq \frac{n -2}{n + 2} \; \frac{1 + \alpha_k^2\frac{n + 2}{4 (n-2)} \frac{\omega_{\text{eff}}^2}{k^2} \frac{2  \left[\text{Re}(b_{0k})\cdot \cos(\alpha_k  z) - \text{Im}(b_{0k})\cdot \sin(\alpha_k  z)\right]}{ \Psi_{0k}  \cos \left(\frac{2 k}{\omega_{\text{eff}}}c_{\text{eff}}z+\Delta_{0k}\right)}}{1 + \frac{2  \left[\text{Re}(b_{0k})\cdot \cos(\alpha_k  z) - \text{Im}(b_{0k})\cdot \sin(\alpha_k  z)\right]}{ \Psi_{0k}  \cos \left(\frac{2 k}{\omega_{\text{eff}}}c_{\text{eff}}z+\Delta_{0k}\right)}}  \;.\; \label{Ceffcorr}
 \end{eqnarray}
 \end{widetext}
where we have considered only the $m=0$ mode in (\ref{analyticPsi})
since the $m\neq 0$ modes vanish when taking the average. 

This expression depends not only on $\alpha_k$ but also on
the initial conditions given by $b_{0k}$ and $\Psi_{0k}$. However, 
from (\ref{psiorder}) we know that $\vert b_{0k}/\Psi_{0k}\vert\simeq \Od(k^2/\omega_{eff}^2)$.
On the  other hand  we have to determine the size of  the parameter $\alpha_k$. With this purpose, we will follow the discussion made in \cite{Arscott} writing $Q(z)$ and a possible solution 
$\delta\phi^{(1)}_k$ as Fourier expansions:
\begin{eqnarray}
\partial_z^2 \delta\phi_k &+& \left( \epsilon_k^2 + \theta_0 + \sum_{r = 1}^{\infty} 2 \theta_r \cos \left(2 r z\right)\right)\delta\phi_k =0\;,\label{EqFloq}\nonumber \\ 
\\
\delta\phi^{(1)}_{k} &=& e^{i \alpha_k z} \sum_{m = - \infty}^{\infty} c_{(2 m) k} e^{i 2 m z}\;.
\end{eqnarray}
We introduce the tentative solution $\delta\phi^{(1)}_k$ in (\ref{EqFloq}),  resulting the following system:
\begin{eqnarray}
c_{(2m) k} + \frac{1}{\theta_0 + \epsilon_k^2 - \left( \alpha_k + 2 m\right)^2} \sum_{l = - \infty }^{\infty} \theta_{2 l} \;c_{(2m + 2l)k} =0&,& \nonumber \\
\\
\text{where } l\neq 0 \text{ and } m = \dots, -1, 0, 1, \dots \; &.& \nonumber
\end{eqnarray}
We eliminate $c_{(2 m) k}$, obtaining the determinant $\Delta_k(\alpha_k)$ and an equation for $\alpha_k$, $\Delta_k(\alpha_k) = 0$. This equation can be rewritten as
\begin{eqnarray}
\cosh \left( i \pi \alpha_k \right) = 1 - 2 \Delta_k(0) \sin\left(\frac{\pi}{2}\sqrt{\theta_0 + \epsilon_k^2}\right)\;. \label{EqDelta}
\end{eqnarray} 
We know the solution when $\epsilon_k = 0$, where one of the modes is periodic. Consequently $\epsilon_k \rightarrow 0 \Rightarrow \alpha_k \rightarrow 0$. Notice that we have also checked this behaviour numerically in Fig. \ref{FigFloquet}. Expanding (\ref{EqDelta}) in $\alpha_k$,
\begin{eqnarray}
\frac{\pi^2 \alpha_k^2}{2} + \dots = 2 \Delta_k(0) \sin\left(\frac{\pi}{2}\sqrt{\theta_0 + \epsilon_k^2}\right)\;. \label{EqDelta2}
\end{eqnarray}
Even if we can not compute $\Delta_k(0)$ analytically  for $\epsilon_k \neq 0$, we know that expanding in $\epsilon_k$ the first correction of the matrix elements is $\mathcal{O}\left(\epsilon_k^2\right)$. Thus from  equation (\ref{EqDelta2}), we expect that $\alpha_k \sim \mathcal{O}(\epsilon_k)$.

Therefore, the correction to the effective sound speed given by (\ref{Ceffcorr}) is:
\begin{eqnarray}
c_{\text{eff}}^2 =\frac{n -2}{n + 2}\left(1+ \Od\left(\frac{k^2}{\omega_{\text{eff}}^2}\right)\right)
\end{eqnarray}
 which is a generalization of (\ref{Csk}) for $n\neq 2$.
\begin{figure*}[t]
\includegraphics[width=\textwidth]{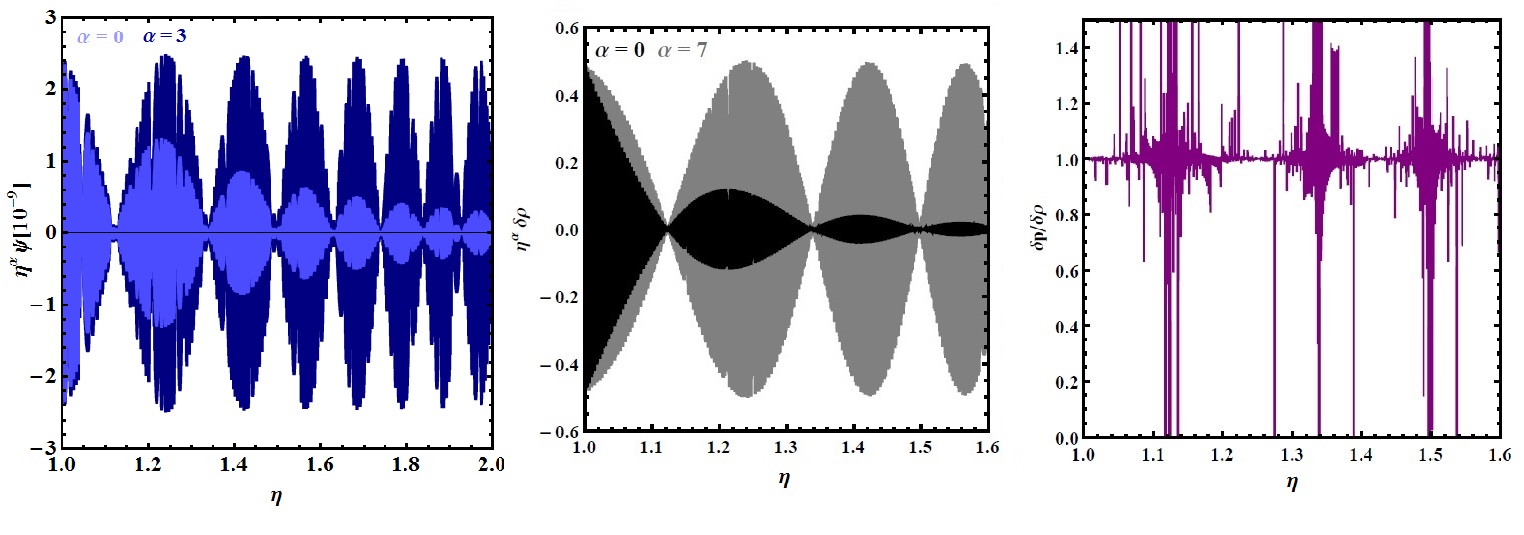}
\caption{\footnotesize{Sub-Hubble mode of a massive scalar field with $k \gg m a$. Those graphics show the numerical solution to the non averaged equations (\ref{Eij}$-$\ref{E0i}), with $V(\phi)= m^2 \phi^2/2$. We work in the same units of Fig. 1. In order to calculate a representative mode, we have set $m = 10$, $k=10^4$, $\delta\phi_k(\eta=1)= 10^{-5}$, $\dot{\delta\phi}_k(\eta=1) = -10^{-5}$, $a(\eta=1)= 1$. The first plot on the left shows clearly the approximation for the scalar perturbation of the metric (blue) made in (\ref{SubPsi}). The perturbation of the energy density (gray) oscillates around zero with an amplitude that decays as $\eta^{-7} \sim a^{-\frac{7}{2}}$.
The last plot shows that the ratio $\delta p_k / \delta \rho_k$ (purple) oscillates around $1$.}}
\label{Figlargek}
\end{figure*}

\subsection{High-$k$ modes}

We have seen that for $ma\gg k$ an oscillating scalar field with power-law
potential behaves as a perfect fluid for which $c_{\text{eff}}^2=\omega$. Let us now consider the opposite limit with $k \gg m a$. For sub-Hubble modes $\Psi_k$ is well approximated by
\begin{eqnarray}
\Psi_k = - 4 \pi G \left(\frac{\dot{\delta \phi}_k\dot{\phi}}{2 k^2} + \frac{V'(\phi)}{2 k^2} \delta\phi_k \right)\;.
\end{eqnarray}
In this case the field perturbation oscillates much faster than the background field. The typical frequency of  $\delta \phi_k$ oscillations would be $k$ \cite{Magnus}, thus
\begin{eqnarray}
\Psi_k \simeq - 4 \pi G \frac{\dot{\delta \phi}_k \dot{\phi}}{2 k^2}\;.\label{SubPsi}
\end{eqnarray}
We can also reach an equivalent expression using (\ref{analyticPsi}). In this limit, we can neglect $V''(\phi)$ from Eq. (\ref{EcpertFlo}) in comparison with $4 k^2/\omega_{\text{eff}}^2$. It implies $\alpha_k \simeq 2 k/\omega_{\text{eff}}\;$. If we assume that the lowest $m$ coefficients are responsible of the main contribution to the Fourier expansion 
\begin{eqnarray}
 c_{m k}\simeq\frac{m+2}{m}\;,\; \text{for}\; \vert m \vert \ll \alpha_k\;.
\end{eqnarray}
And, thus,
\begin{eqnarray}
&& \Psi_k \simeq \Psi_{k,part} \simeq - \sum_{\vert m \vert \ll \alpha_k} \frac{m + 2}{m} \left(\text{Re}\left(b_{mk}\right) \right.
\\
&& \left.\cos\left((2m+\alpha_k)\eta\right)- \text{Im}\left(b_{mk}\right) \sin\left((2m+\alpha_k)\eta\right)\right)\;.\nonumber
\end{eqnarray}

The gravitational potential, and accordingly the density perturbation, oscillates around zero
as shown in Fig.  \ref{Figlargek}. Because of this fact, all the perturbations vanish in average but the effective sound speed is $c_{\text{eff}}^2 = 1$ (see Fig. \ref{Figlargek}) according to (\ref{Cseff}), since in both $\delta p_k$ and $\delta \rho_k$, the kinetic term dominates.

\vspace{0.2cm}

\section{Conclusions}
In this work we have shown that a coherent homogeneous scalar field oscillating in 
a power-law potential behaves as an adiabatic perfect fluid with 
constant equation of state 
both at the background and perturbation levels. Thus, scalar perturbations 
are shown to propagate, to the leading order in $k/\omega_{\text{eff}}$, with a sound speed given by 
$c_{\text{eff}}^2 = \omega=(n-2)/(n+2)$. The first correction to this expression 
is shown to be $\Od\left(k^2/\omega_{\text{eff}}^2\right)$.
The robustness of the result has been shown by studying the exact system in the sub-Hubble and super-Hubble limits as well as in the numerical computations.

These results extend previous analysis done in the massive case $n=2$ and
opens the possibility of using this kind of models as perfect fluid analogues 
in different cosmological contexts. In the general case, we have shown that 
there are departures from the perfect fluid 
behaviour on small scales with a cut-off around $k\simeq \omega_{\text{eff}}$ very much as in the harmonic case.  Notice that 
for $n<2$ the negative value of $c^2_{\text{eff}}$ suggests the generation of instabilities as found in previous works 
in the homogeneous case.   

The analysis performed in this paper could be extended to higher-spin oscillating fields.
In particular in the massive vector case, which as shown in \cite{Cembranos:2012kk} behaves as non-relativistic
matter at the background level, this study would allow to determine the growth of
structures and its viability as dark matter candidate. Work is in progress in 
this direction \cite{IP}.

{\bf Acknowledgements:}
This work has been supported by MICINN (Spain) project numbers FIS2011-23000, FPA2011-27853-01, FIS2014-52837-P and Consolider-Ingenio MULTIDARK CSD2009-00064. S.J.N.J. acknowledge support from  Complutense University under grant CT4/14. We would like to thank Alberto Diez Tejedor for his useful comments.

\end{document}